\documentclass{aa}  

\usepackage{natbib}
\usepackage{graphicx}
\usepackage{txfonts}
\usepackage{hyperref}
\usepackage{lscape}
\usepackage{ amssymb }
\usepackage{color,soul}

\newcommand\e{\'{e}}
\newcommand\s{$\sim$}
\newcommand{\angstrom}{\textup{\AA}}

\begin{document} 

  \title{Color gradients reflect an inside-out growth in early-type galaxies of the cluster MACS J1206.2-0847}
  \subtitle{}
  \author{
  V. Marian\inst{1}\fnmsep
  \thanks{Current adress: \newline
  Max Planck Institut for Astronomy, 
  K\"onigstuhl 17, 
  D-69117 Heidelberg
  }
  \and
  B. Ziegler\inst{1}
  \and
  U. Kuchner\inst{1,2}
  \and
  M. Verdugo\inst{1}
         }

  \institute{University of Vienna, Department of Astrophysics, 
              T\"urkenschanzstrasse 17, A-1180 Vienna
  \and  School of Physics \& Astronomy, The University of Nottingham, University Park, Nottingham, NG7 2RD, UK\\
             \email{marian@mpia.de}}

  \date{}

  \abstract
   {}
   { 
    Color gradients of galaxies are a powerful tool for resolving the variations of stellar populations within galaxies. We use this approach to explore the evolution of early-type galaxies in the core of the massive galaxy cluster MACS J1206.2-0847 at z = 0.44.
   }
   {
   We used imaging data in 12 filters (covering a wavelength range from $400 - 1600$nm) from the Hubble Space Telescope provided by the CLASH survey, as well as additional spectral information from its follow-up program, CLASH-VLT.
   We performed multiwavelength optimized model fitting using Galapagos-2 from the MegaMorph project to measure their photometric parameters (total integrated magnitudes, effective radii $r_e$ , and S\e rsic indices $n$). We used them to derive color gradients for the colors $g_{475} - I_{814}$, $r_{625} - Y_{105}$, $I_{814} - H_{160}$ , and $Y_{105} - H_{160}$ at radii ranging between 0.1 - 2 $r_e$ for 79 early-type cluster galaxies. 
   From synthetic spectral models that use simple star formation recipes, we inferred ages and metallicities of the stellar population at different locations within each galaxy and characterized their influence on the radial color trends. 
    }
   {
   Early-type galaxies show a substantial decrease in effective radii $r_e$ with wavelength. We measure that galaxy sizes are \s 25~\% smaller in the red $H_{160}$ filter than in the blue $r_{625}$ filter but maintain a constant (within 3$\sigma$) S\e rsic index $n$ with wavelength. 
   We find negative color gradients in all colors with slopes ranging between -0.07 and -0.17 mag dex$^{-1}$ and with no obvious dependence on total magnitude, stellar mass, or location inside the cluster core. 
   We explain the observed radial trends of color gradients as a result of the ages and metallicities of the respective stellar populations. 
   Red galaxy cores are typically \s 3 Gyr older and more enriched in metals than the galaxy outskirts, which are of solar metallicity.
   }
   {
   Our results support the predictions from hydrodynamical cosmological simulations, which describe a passive evolution combined with an inside-out-growth of early-type galaxies. In this scenario, galaxies assemble their stellar mass primarily in the outskirts through the accretion of mass-poor satellites and thus manifest the observed trends of color-, metallicity- and age gradients.
   }


\titlerunning{Color-gradients reflect an inside-out growth in early-type cluster galaxies}
\maketitle

\section{Introduction}

It is well known that the bulk of galaxies at redshifts lower than z~\s~1 can be described by a bimodal distribution; a notion which has initially been introduced by Edwin Hubble \citep{hubble_26_distance_measurements}. The crude classification scheme that separates galaxies into disk-dominated, blue, and spiral system galaxies on the one hand and spheroid-dominated, red, and elliptical galaxies on the other
hand is likely too superficial. The scheme has since been updated several times: lenticular S0 galaxies and irregular galaxies have been added \citep[e.g.,][]{kormendy_12_revised_classification}. 
Nevertheless, observations continue to confirm that numerous physical properties of galaxies, such as star formation rate, gas content, stellar mass, and metallicity, roughly correlate with galaxy morphologies in a bimodal way \citep{conselice_06_fundamental_props_gal}. 

The morphology of a galaxy is the product of its formation and evolution over time, including internal perturbations or interactions with the surrounding intergalactic or intracluster medium, as well as with other galaxies.
Early-type galaxies represent the majority of galaxies at local and intermediate redshifts, and contain \s 60~\% of all stellar mass (\citet{hogg_02_luminosity_density_red_gal}; \citet{driver_06_mill_gal_cat_morph_class_bimodality}). Following the \textit{\textup{morphology-density -- relation}} they dominate the highest density regimes (\citet{dressler_80_morpho_density}; \citet{dressler_84_evo_of_gal_clusters}), such as cluster centers. An understanding of the evolution of this morphological class of galaxies is therefore essential for discerning and explaining the general development of galaxies.

Over the course of the past decades, several frameworks for the formation and evolution of galaxies were introduced, revised, and sometimes even rejected again. Two of the most prominent are the monolithic or top-down model and the hierarchical or bottom-up scenario. The former predicts the formation of spheroids as a result of a global starburst at very early epochs of the Universe, followed by a passive evolution to the present day. Spheroids may subsequently form disk components by accreting gas from their surroundings. Because the gas content is limited, this leads to distinct epochs of star formation, thus requiring spheroids to precede disks in the monolithic scenario \citep{eggen_62_monolithic, larson_74_monolithic}. Conversely, the top-down scheme envisions large spheroids as the results of merging events of two disk galaxies, thus requiring disks to predate spheroids. If the progenitors are gas rich, the merging events lead to the disruption of the disk as well as intense bursts of star formation, which thus form the bulk of the newly created elliptical galaxies \citep{renzini_06_stellar_pop_diagnostics_E_gal_form}.
Alternatively, the merger of two or more quiescent galaxies ultimately creates a massive, early-type system \citep{bell_04_etg_dry_merger}.

However, some observations contradicted these two theoretical scenarios. For example, it has been shown that more massive galaxies form their stars at earlier times than lower-mass galaxies (the galaxy down-sizing; \citep{bender_96_bodo_ref, cowie_96_downsizing, bundy_06_downsizing}), which apparently contradicts the hierarchical scheme.

As a result of these complications, a new, `inside-out-growth' scheme arose and became widely accepted \citep[e.g.,][]{carrasco_10_inside_out, hopkins_10_inside_out_sim, vandokkum_10_growth_massive_gal_z_2}. In this inside-out scenario, compact early-type galaxies grow through `dry' minor mergers, which increase the galaxies in size, but hardly in mass. This picture is consistent with the description of a change in the cosmic star formation history at $z \sim$ 1.7 from a hot-mode evolution to a cold-mode evolution \citep{driver_13_Two_phase_galaxy_evo_cosmic_SFH}. The first mode is characterized by the formation and growth of spheroids through mergers and/or collapse, and the latter describes the formation and growth of disks by gas infall and minor mergers.

Numerous studies have corroborated this model through observations of the early formation of  early-type galaxy cores and the growth in size, which favors a scenario of growth through minor mergers \citep{buitrago_17_grow_by_merger, daddi_05_passive_evo_etgs, hilz_13_inside_out, mclure_13_size_growth, newman_12_size_growth,
trujillo_07_strong_size_evo_most_massive_gal_since_z_2,
trujillo_11_size_evo_etg_since_z_1_puffing_up_vs_minor_merger, vandesande_13_inside_out_growth, 
vandokkum_08_compactness_etgs_no_monolithic,
vandokkum_10_growth_massive_gal_z_2, vandokkum_14_high_z_dense_cores} , and galaxies evolving from $z \sim$ 1.5 to $z = $ 0, while increasing their effective radius by a factor of \s\ 1.5 
\citep{buitrago_08_size_evo_massive_gal,
longhetti_07_kormendy_rel_massive_etgs_evidence_size_evo} and a factor of \s\ 4 since $z \sim$ 2 \citep{chan_16_sizes_colorgrads_etc_in_massive_cluster}, which relates to $r \varpropto (1+z)^{-1.48}$ \citep{van_der_wel_14_3dhst_candels_evolution_size_mass_distr}. 
The majority of stars in these systems was formed at high redshifts ($z \approx$ 3-5) and on short timescales ($\tau \sim$ 1 Gyr) \citep{thomas_05_epochs_etg_form_func_environment, thomas_09_dm_scaling_rel_assembly_epoch_coma_etgs}. This can be derived from correlations between global properties, such as the color-magnitude relation, the fundamental plane \citep{bender_92_dyn_hot_gal_struct_prop}, or the Mg-$\sigma$ relation \citep{ziegler_97_mg_sig_rel}.
The lack of small but massive galaxies in the local Universe suggests that they are the progenitors of the local massive early-type population \citep{patel_13_inside_out_progenitors, williams_14_progenitors}.

This model can also be explained successfully by hydrodynamical cosmological simulations \citep{hopkins_10_inside_out_sim}. \citet{naab_09_minor_merger_and_size_evo_etgs} have shown that while the accretion of stripped infalling stellar material increases the size of elliptical galaxies, the central concentration is reduced by dynamical friction. Using the virial theorem, they demonstrated that minor mergers increase the radius of a galaxy as the square of the mass, but not the mass to the same extent. In this way, compact high-redshift spheroidal galaxies can evolve to representative sizes and concentrations of local elliptical galaxies \citep{bezanson_09_virial_formula_growth}.

As a result, individual evolutionary paths of galaxies are reflected by the distributions of their stellar populations within.
In order to probe galaxy evolution scenarios, we therefore correlate age-, metallicity-, and the resulting color-gradients, that is, the age, metallicity, and color dependence on radius, to the underlying spatial distribution of the stellar population. For this, we examine galaxy light profiles as a function of radius in different passbands.

In the classic monolithic dissipative model, metal-rich gas flows inward during the initial gravitational collapse, leading to higher metallicities in the center. This is characterized by negative metallicity gradients.
In addition, the deep gravitational potential well in the center causes gas to continue flowing inward, which triggers star formation and enriches the gas further. 
Because this process lasts longest in the center, the monolithic model predicts null or positive age gradients. 
Metallicity gradients dominate the color profiles, which leads to strong negative color gradients of values greater than -1 mag $\mathrm{dex^{-1}}$ in radius for the classical, and -0.5 to -0.3 mag $\mathrm{dex^{-1}}$ for the revised monolithic model \citep[and references therein]{montes_14_age_and_Z_gradients_m87}.

On the other hand, the hierarchical dry merger, which predicts a mix of stellar populations, and inside-out-growth scenarios, which add new stellar content as an envelope to the existing galaxy, lead to much shallower metallicity and color gradients. The reason is that pre-existing gradients are diluted as a result of the merger(s). 

In support of the current scenario of galaxy evolution, several studies at all redshifts describe a significant variation in recovered effective radius $r_e$ of galaxies as a function of wavelength, suggesting the existence of different stellar populations in individual galaxies. \citet{la_barbera_02_opt_and_struct_prop} examined cluster early-type galaxies with $n \gtrsim$ 4 at intermediate redshift ($z$ = 0.31) and found  that effective radii are \s\ 40~\% smaller in the near-infrared than in the optical. This was confirmed by \citet{la_barbera_10_gal_params_grizYJHK} and \citet{kelvin_12_gama_struc_investigation_via_model}, whose results show that on average, early-type galaxies are more concentrated at longer wavelengths. 
The S\e rsic index $n$ of spheroidal galaxies (i.e., high $n$ galaxies) depends only weakly on wavelength: their $n$ remains relatively stable at all wavelengths, with slightly lower values at shorter wavelengths \citep{kelvin_12_gama_struc_investigation_via_model, vulcani_14_gama_sizes_and_profiles_with_megamorph, kennedy_15_gama_wavelength_dependence_structure}.

Color gradients in local early-type galaxies have been measured for the past 30 years \citep[e.g.,][]{vader_88_photometry_of_etg_part_2}. Most publications report negative color gradients, which correspond to a decrease in metallicity with radius  \citep{peletier_90_ccd_surface_photometry_of_galaxies,peletier_90_nir_photometry_etg, la_barbera_09_origin_color_grads_etg_compactness_high_z} and small positive age gradients, which relates to the observed size growth of galaxies 
\citep{montes_14_age_and_Z_gradients_m87, wu_05_optical_NIR_color_profiles_etg_age_Z_grads}.
Recent studies using spectroscopic data on limited sample sizes confirm these observations \citep{goddard_17_spec_gradient_environment, kuntschner_10_spec_gradients, sanchez_blazquez_07_spec_gradients}.

The advancement in technology allowed expanding the color gradient measurements to more distant galaxies. 
Several investigations of high-redshift passive galaxies confirmed that metallicity is the dominant driver for color gradients in these galaxies. Furthermore, this result seems to be independent of environment \citep{saglia_00_evolution_color_gradients_of_etg, smail_01_phot_study_ages_Z_etg_cluster, tamura_00_color_gradients_etg, tamura_00_origin_color_gradients_etg}. 

\begin{figure*}[!ht]
\centering
\includegraphics[width=18.5cm]{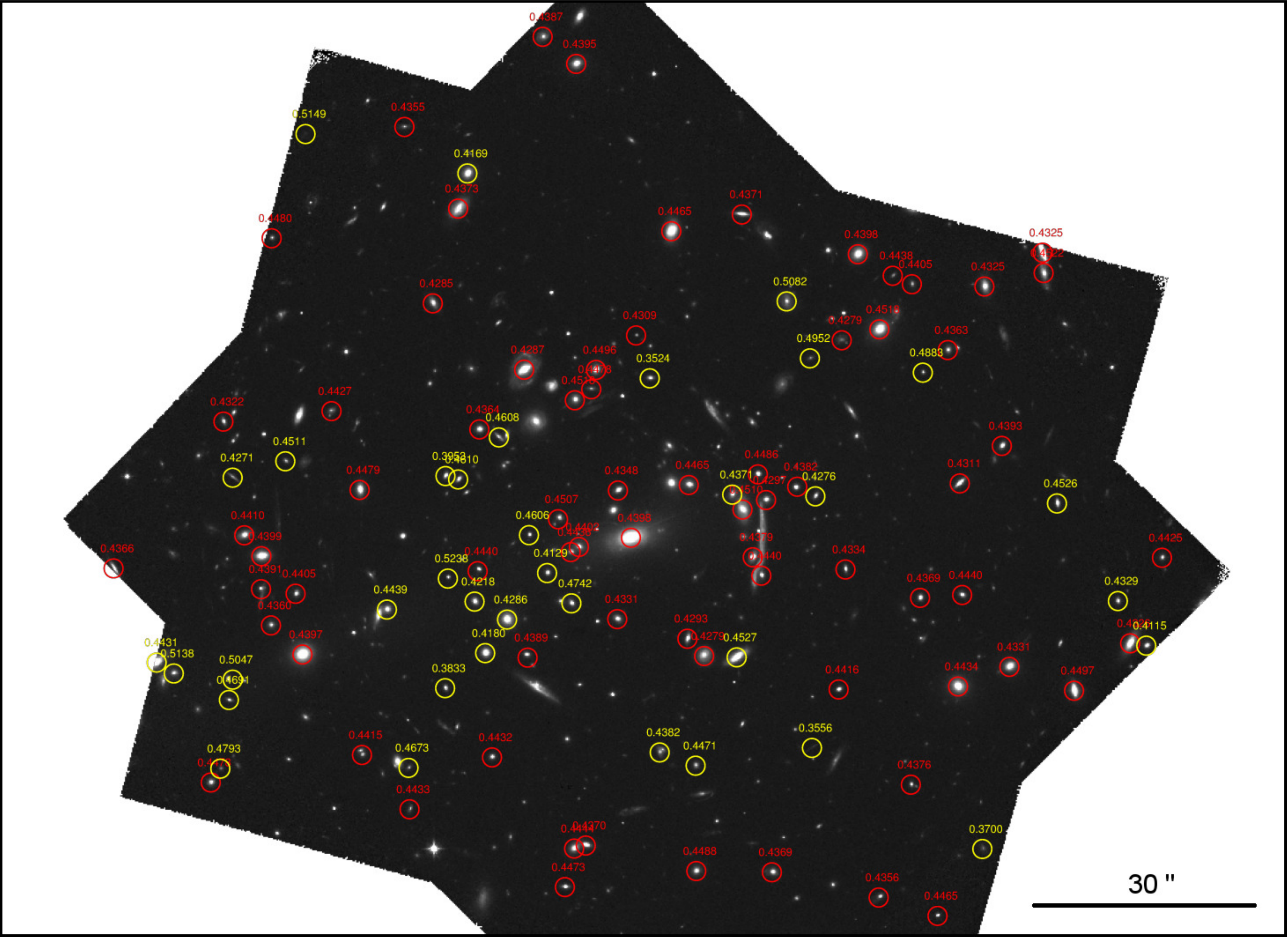}
\caption{Initial sample of early-type galaxies in MACS 1206. Red circles indicate galaxies with spectroscopic redshifts, and yellow circles show galaxies with photometric redshifts. Additionally, we plot the respective redshifts.}
\label{1206_selection}
\end{figure*}

However, color gradients vary systematically with redshift.
Color gradients of elliptical galaxies at redshifts $1 < z < 2$ appear significantly steeper, reaching twice the values of the local elliptical population \citep{gargiulo_11_colour_gradients_etg,gargiulo_12_colour_and_stellar_prop_etg,guo_11_color_and_stellar_pop_gradients}. This implies an important contribution by an age gradient at high redshifts \citep{chan_16_sizes_colorgrads_etc_in_massive_cluster}, when galaxies enter their phase of passive evolution \citep{de_popris_15_morphological_evo_red_sequence}. Over the following \s 8-9 Gyr, the stellar population throughout a galaxy evolves from being dominated by young, blue stars to the evolved old, red, and remaining long-living stars without any new star formation. The age gradient thus tends to null for galaxies at $z \sim 0$.

The evolution of color gradients as well as galaxy sizes over cosmic time, and their observed dependence on wavelength, give reason to 
endorse the inside-out-growth model. This is also in agreement with chemodynamical simulations, which succeed in reproducing the observed variety of metallicity gradients in elliptical galaxies \citep{kobayashi_04_chemodyn_sim_etg_evo_Z_grads}.

The combination of our comprehensive, large, and deep high-quality HST sample of cluster early-type galaxies, observed in 16 different filters, and our simultaneous multi-band fitting approach allows us to extensively verify the current evolution model of early-type galaxies. Since all sources lie within one cluster, we are also able to additionally assess a possible dependence of the color gradients on environment or stellar mass.

We examine a sample of 79 carefully selected early-type galaxies in the cluster MACS J1206.2-0847 (MACS 1206 in the following) observed by the HST multi-cycle treasury program "The Cluster Lensing And Supernova survey with Hubble" \citep[CLASH\footnote{\url{https://archive.stsci.edu/prepds/clash/}}, ][]{postman_12_clash}. Initially detected by the MAssive Cluster Survey \citep[MACS,][]{ebeling_01_macs_survey}, MACS 1206 is one of the 25 massive galaxy clusters of CLASH, observed in 16 filters, ranging from the near-UV to the near-IR. 
We analyze color gradients of our sample galaxies for the colors $g_{475} - I_{814}$, $r_{625} - Y_{105}$, $I_{814} - H_{160}$ , and $Y_{105} - H_{160}$, as well as the variations in structural parameter S\e rsic index $n$ and effective radius $r_e$ over the observed wavelength range. 
To constrain ages and metallicities of the stellar population at different locations within each galaxy, we compare measured colors with colors predicted by simple stellar population models. This allows a link of age and metallicity variances with the observed color gradients.

The paper is structured as follows: In Sect.~\ref{Data} we describe our data sample, the global properties of MACS 1206, and our selection of targets. In Sect.~\ref{Sec3} we illustrate the modeling of our sample using \texttt{Galapagos-2}, a software developed by the MegaMorph project, and our final selection of early-type galaxies. In Sect.~\ref{sec4} we present our results of structural parameter effective radius $r_e$ and S\e rsic index $n$, and we explain the derivation of color gradients, while in Sect.~\ref{sec5} we compare our results to those provided by stellar population models. Finally, in Sect.~\ref{sec6} we discuss our results, and we summarize in Sect.~\ref{sec7}.

We adopt the concordance cosmology with $H_0$ = 70 km s$^{-1}$ Mpc~$^{-1}$, $\Omega_{\Lambda}$ = 0.7, and $\Omega_M$ = 0.3 \citep{spergel_03_wmap_concordance_cosmology}, in agreement with other studies examining the \textit{CLASH} clusters \citep[e.g., ][]{ annunziatella_15_clash-vlt_stellar_mass_function_mass_density_profile_1206, balestra_clash_vlt_0416, biviano_13_clash_vlt_1206, girardi_15_clash_vlt_substructure_1206_from_kinematics, jouvel_14_clash_phot_z,kuchner_17_paper, maier_16_1206, ogrean_15_chandra_jvla_0416, presotto_14_intracluster_light_1206, zitrin_12_clash_inner_mass_profile_1206}.

\section{Data}
\label{Data}

MACS 1206 is a massive ($L_X \sim 2.4 \times 10^{45}$ erg s $^{-1}$) cluster at z $\sim$ 0.44. According to X-ray observations, the cluster appears to be in a  relaxed state. 
The central brightest cluster galaxy (BCG) at RA$_{2000} = 12^{h}06^{m}12^{s}.15$ and Dec$_{2000} = -8^{\circ}48'3.''48$ is located at the peak of the X-ray emission and the gravitational  mass center \citep{umetsu_12_mass_distribution_1206}. 
The mass profiles derived from dynamical and strong- and weak-lensing methods \citep{zitrin_12_clash_inner_mass_profile_1206} are in excellent agreement, with a virial mass of $M_{200} \sim 1.4 \times 10^{15}M_{\odot}$ and a virial radius of $r_{200} = 1.98$ Mpc \citep{biviano_13_clash_vlt_1206}. 

The cluster also exhibits a significant WNW-ESE elongated intracluster light (ICL) component, which indicates that despite the overall relaxed appearance \citep{eichner_13_halo_truncation_1206}, interactions between galaxies are still ongoing and result in tidal disruptions that feed the ICL. 
This is supported by the finding that the ratio of giant ($M_*/M_{\odot} > 10^{10.5}$) to subgiant galaxies ($10^{9.5} < M_*/M_{\odot} < 10^{10.5}$) is highest in the innermost regions of the cluster ($R > 0.5$ Mpc). The ratio drops to its minimum in the adjoining region at distances between 0.5 and 1 Mpc away from the BCG \citep{annunziatella_15_clash-vlt_stellar_mass_function_mass_density_profile_1206}. 
In addition, an enhancement of red galaxies with strong H$\delta$ absorption, indicating a recent star formation event 1-2~Gyr ago, is found in the center and along the ICL. This further implies recent or even ongoing interactions on galaxy scales \citep{girardi_15_clash_vlt_substructure_1206_from_kinematics,mercurio_15_lum_funct_1206,presotto_14_intracluster_light_1206}.

The images provided by the CLASH pipeline are already cleaned, processed, aligned, and co-added using the MosaicDrizzle software \citep{koekemoer_11_drizzle_HST_mosaics}, which corrects for cosmic-ray contamination and compensates for shifts and rotations. 
Precise and reliable photometric redshift measurements were derived for 16 filters \citep{jouvel_14_clash_phot_z}, which we used in our selection process.  

Complementary to the photometry, we also used results from the spectroscopic follow-up program CLASH-VLT \citep{rosati_14_clash_vlt_dm_distr_messenger}, which observed 13 out of the 25 CLASH clusters with the low-resolution blue and medium-resolution grisms of the VLT VIMOS instrument. 
Analyzing the locations of the sources in the projected phase space, \citet{biviano_13_clash_vlt_1206} identified $\sim$ 600 confirmed member galaxies in projected phase space of MACS 1206.
Together with photometric redshifts using archival images obtained with the Suprime-Cam instrument on the Subaru telescope, they acquired a redshift range of 0.34 $< z <$ 0.54 for galaxies to be defined as cluster members \citep[see also][]{annunziatella_15_clash-vlt_stellar_mass_function_mass_density_profile_1206}

Combining these membership determinations with the photometric redshift determinations by \citet{jouvel_14_clash_phot_z}, we obtain an inital sample of 110 galaxies for the ACS $I_{814}$ band, which serves as our reference band in the ensuing analysis. Out of this sample, 74 galaxies possess a spectroscopically derived redshift, while 36 have photometric redshift determinations. As the field of view of ACS is larger than that of WFC3/IR, we only include sources whose positions are covered in all filters to optimize the accuracy of the output of the modeling process. As a consequence, our galaxies are located well within the innermost regions of the cluster, and the largest centric distances are less than half of the virial radius. Nevertheless, the galaxies appear evenly distributed (Fig. \ref{1206_selection}). This should avoid any biases such as contamination by the ICL, and orientation or location in the cluster. At the cluster redshift, one arcsecond corresponds to $\sim$ 5.67 kpc.

\section{Modeling of galaxies}
\label{Sec3}

To determine the structural parameter S\e rsic index $n$ and effective radius $r_e$ needed for the subsequent analysis, we simultaneously modeled our sample galaxies in 12 optical and near-infrared HST observations.
\footnote{Because of the low S/N in the WFC3/UVIS bands, we discarded the observations in these four filters in our model fittings.}
For this, we used Galapagos-2, a tool provided by the \texttt{MegaMorph} project \citep{bamford_12_sed_for_gal_comp, haeussler_13_megamorph}, which is a wrapper for \texttt{Source Extractor} \citep{bertin_96_sextractor} and \texttt{GALFIT-M}, an extension of the widely used galaxy model fitting tool \texttt{GALFIT} \citep{peng_02_galfit,peng_10_galfit_3}. \texttt{GALFIT-M} is capable of simultaneously fitting intensity profiles to galaxies with multi-band data, which has been shown to enhance the stability of the fitting results as well as increase the signal-to-noise ratio (S/N), that is, lowers the magnitude limit in comparison to single-band fits \citep[e.g.,][]{haeussler_13_megamorph}.

The main upgrading of \texttt{GALFIT-M} relates to the replacement of the initial model parameter outputs of \texttt{GALFIT} with wavelength-dependent Chebyshev polynomials, where the coefficients are the fitted parameter results in the respective passbands. 
The degree of freedom is defined by the user and ranges from constant over all wavelengths to completely independent. 
Hence, the user is free to decide to which extent the values for each fitted parameter are connected to each other.

Since the focus of this study lies on early-type galaxies, we chose to model all galaxies with a one-component fit \footnote{Testing for a subset of our sample more elaborate two-component fits in regard of computational time showed no significant difference in the results.}, that is, with a free single S\e rsic profile. 
For early-type galaxies, this choice is well established, and it has been shown that the shape of elliptical galaxies is well represented by a single S\e rsic profile, often averaging at $n \sim 4$ \citep[e.g., ][]{deVaucouleurs_48_profile, haeussler_13_megamorph}. 

We chose to fix the central X and Y positions of the galaxies, their axial ratios, and position angles to be constant over all wavelengths. 
In this way, we avoided artificial color gradients, which may be introduced through different shapes and/or orientation. 
The structural parameter effective radius $r_e$ and S\e rsic index $n$ are assumed to be best described by a second-order polynomial, although a linear function is expected to be sufficient for the brightest elliptical galaxies \citep{haeussler_13_megamorph, kuchner_17_paper}. 
Following conventions, we constrained $n$ to values between 0.2 and 8, and defined the maximum limit of the radius $r_e$ as 400 pixels, which corresponds to $\sim$ 150kpc at z $\sim$ 0.4. 
Model magnitudes were fit freely without constraints, meaning that S\e rsic magnitudes in the different bands can vary independently of the wavelength. 

Of the 110 sample galaxies, 103 successfully returned model values. For the 7 remaining galaxies without successful completion of the fitting process (one is the BCG of the cluster), the maximum time we allowed for the derivation of a model was exceeded. 
A visual inspection showed that these galaxies are embedded in extended light emission either from intracluster light or bright neighboring sources. Galaxies in very crowded regions are fit simultaneously, which further increases the computational time, in some cases exceeding the chosen time-frame. 

\begin{figure}
\centering
\includegraphics[width = 10cm]{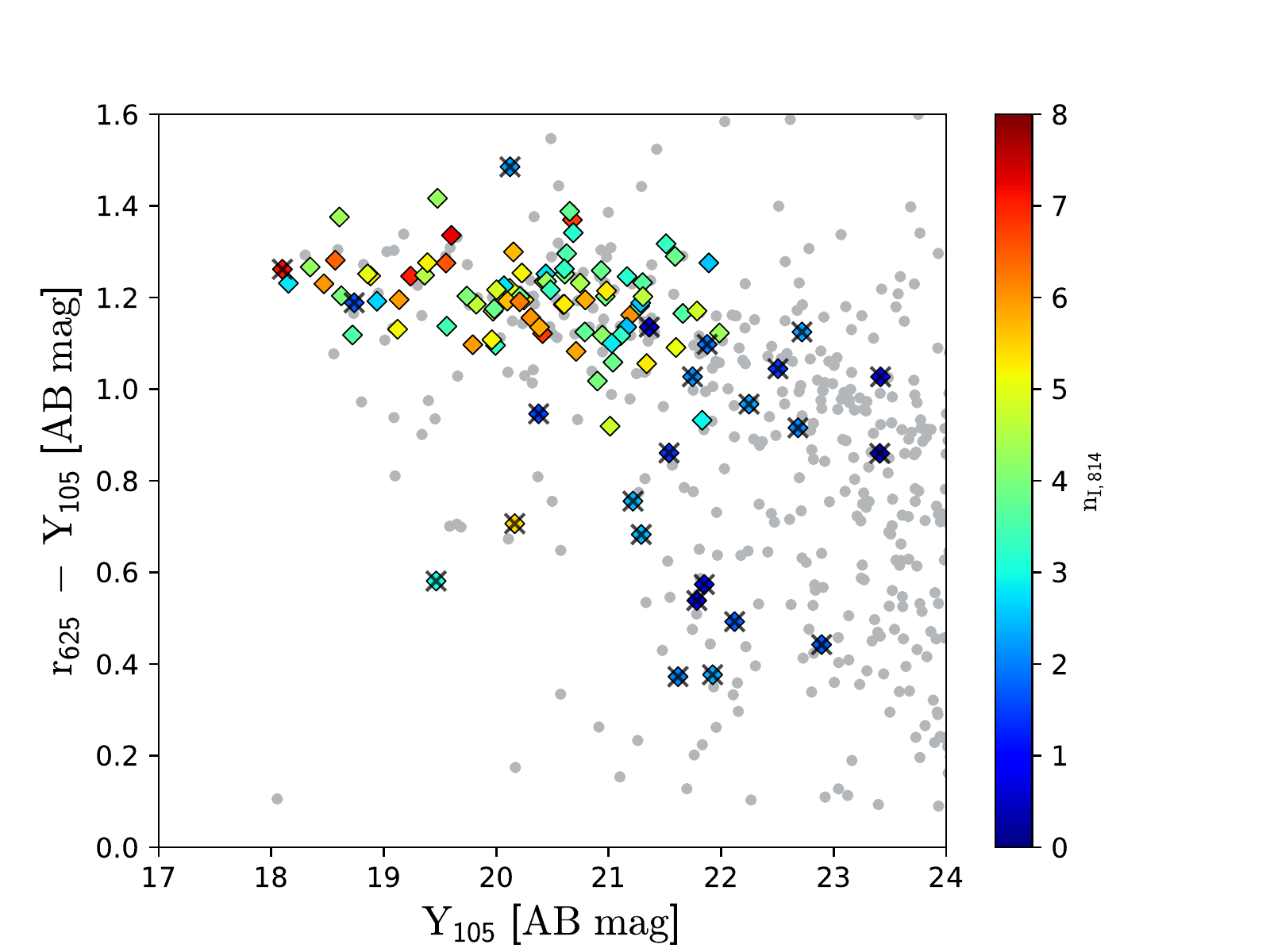}
\caption{Color-magnitude diagram for the fitted galaxies in MACS 1206. The gray dots represent all detected sources in the WFC3/IR field of view, overlaid by the fitted galaxies, color-coded according to their S\e rsic index $n$ in $I_{814}$. Fitted galaxies (colored diamonds) with black crosses denote sources that are not considered in the subsequent analysis due to an $n<2.5$ or morphological features, such as spiral arms or distortions. As expected, our final sample galaxies all lie on the red sequence.}
\label{1206_cmd}
\end{figure}
\begin{figure*}
\centering
\includegraphics[width=19cm]{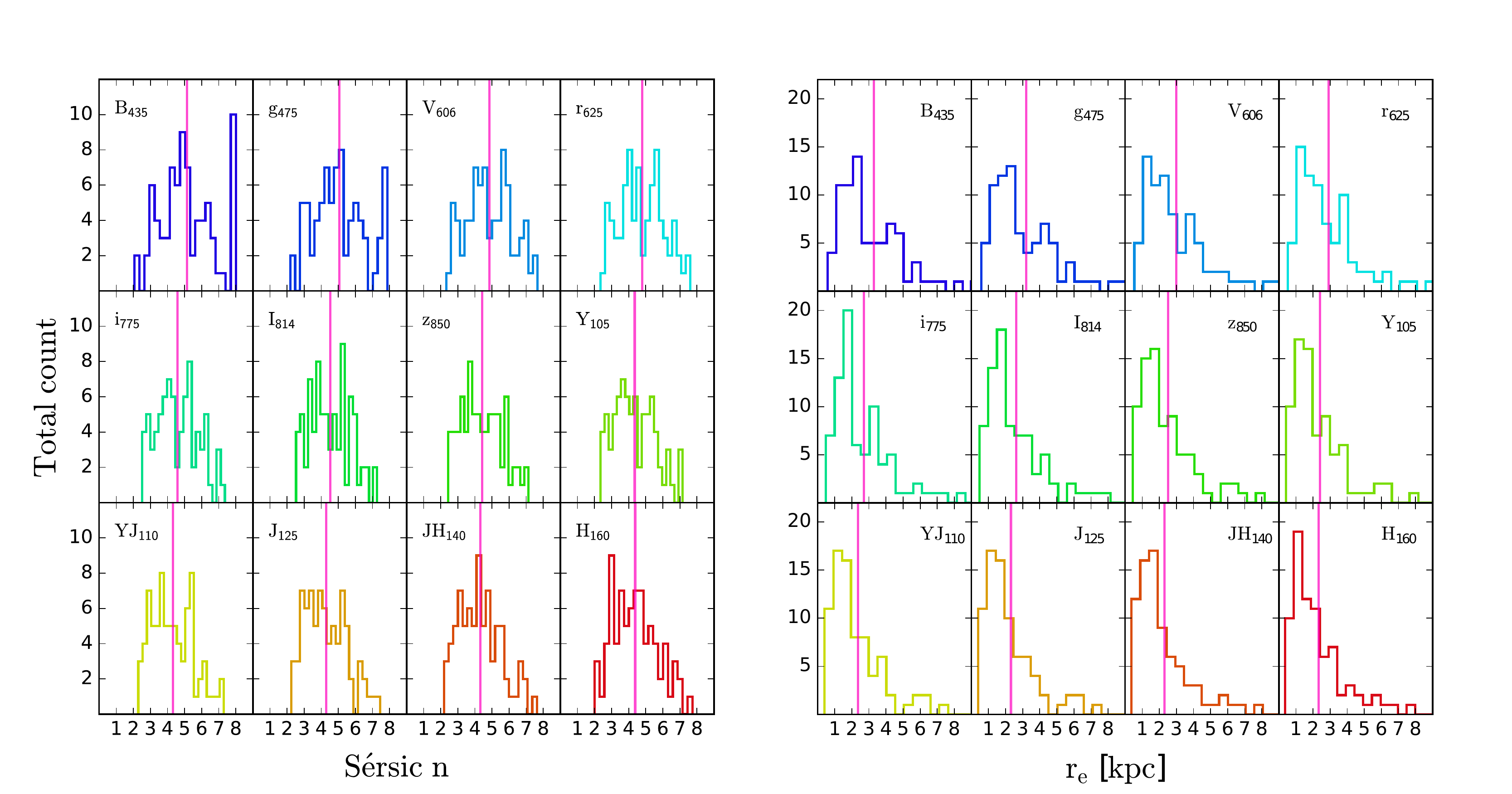}
\caption{Left side: Distribution of S\e rsic index $n$ in the 12 bands for the 79 galaxies contained in the final sample. The magenta line shows the respective mean, and the filter name is given in each panel. In all 12 bands our mean S\e rsic indices show typical values for early-type galaxies. Right side: Analogously to the left side, the distributions for the effective radius $r_e$ are shown. A clear trend to smaller sizes with increasing wavelength is visible.}
\label{1206_histos}
\end{figure*}

Fig. \ref{1206_cmd} shows the color-magnitude diagram of galaxies in MACS 1206, where we plot integrated, modeled AB magnitudes in $Y_{105}$ on the x-axis and $r_{625} - Y_{105}$ colors on the y-axis. The colored points represent the sample galaxies, color-coded from blue to red according to the derived S\e rsic indices $n$ in the $I_{814}$ band, from low to high values.
Gray dots represent all galaxies that are visible in the WFC3/IR field of view, independent of their redshift.
The sample galaxies are mainly located along the red sequence, and most S\e rsic indices fit expectations of early-type galaxies, which dominate the inner regions of galaxy clusters, which is the region we cover here.

However, some galaxies stray from this expectation.
We therefore further separated the galaxies into morphological early- and late-type galaxies according to their S\e rsic index and visual shape.
Following the convention of previous
publications \citep[e.g.,][]{shen_03_size_distribution_galaxies_sdss,barden_05_gems_surface_brightness_evo_disk_galaxies,vulcani_14_gama_sizes_and_profiles_with_megamorph},
we separated galaxies into disk-dominated galaxies when $n < $ 2.5 and spheroid-dominated galaxies when $n >$ 2.5 in $I_{814}$, which has been the reference band throughout the fitting process.
In addition, we visually inspected the sources in the $I_{814}$ band three times, each time shuffling the galaxies to a new random order.
All galaxies with $n < $ 2.5 and those indicating spiral or peculiar features  (independent of their $n$ value) were discarded from further considerations. In Fig. \ref{1206_cmd}, they are marked by a black cross. The majority of the discarded galaxies lies well below the red sequence.
Applying these constraints leads to a final sample of 79 (57 of which are spectroscopically confirmed) carefully selected smooth, shperoid-dominated, red  galaxies in the central region of MACS 1206.

\section{Results}
\label{sec4}

\subsection{\texorpdfstring{Variations in structural parameters $n$ and $r_e$}{Variations of structural parameters n and r\_e}}

In this section we use the measurements of S\e rsic index $n$ and effective radius $r_e$ that are available for each galaxy in 12 bands to analyze their variations with wavelength.

\begin{table*}
\caption{Mean values for $n$ and $r_e$ in kpc for the MACS 1206 sample in each filter. The uncertainties (i.e. standard error of the mean) are derived by using $\sigma/\sqrt{N}$, with $\sigma$ being the standard deviation and $N$ the number of galaxies in our sample.}      
\label{1206_n_re_var_table}      
\centering          
\begin{tabular}{c c c c c c c c c c c c c}     
\hline\hline       
& B$_{435}$ & g$_{475}$ & V$_{606}$ & r$_{625}$ & I$_{775}$ & I$_{814}$ \\ 

\hline                    
 $n$ & 5.15 $\pm$ 0.18 & 5.06 $\pm$ 0.17	& 4.85 $\pm$ 0.15 & 4.79 $\pm$ 0.14 & 4.59 $\pm$ 0.14 & 4.53 $\pm$ 0.14
\\ 
$r_e$ [kpc] & 3.30 $\pm$ 0.23 & 3.20 $\pm$ 0.23 & 2.99 $\pm$ 0.22 & 2.92 $\pm$ 0.22 & 2.70 $\pm$ 0.21 & 2.63 $\pm$ 0.21
 \\ 
\hline  
 &  &  &  &  &  &  \\ 
 & z$_{850}$ & Y$_{105}$ & YJ$_{110}$ & J$_{125}$ & JH$_{140}$ & H$_{160}$ \\ 
\hline
$n$ & 4.42 $\pm$ 0.14 & 4.35 $\pm$ 0.14 & 4.31 $\pm$ 0.14 & 4.30 $\pm$ 0.14 & 4.32 $\pm$ 0.14 & 4.38 $\pm$ 0.15 \\ 
$r_e$ [kpc] & 2.51 $\pm$ 0.20 & 2.42 $\pm$ 0.20 & 2.36 $\pm$ 0.20 & 2.32 $\pm$ 0.20 &	2.30 $\pm$ 0.20 & 2.32 $\pm$ 0.20
 \\
\hline
\end{tabular}
\end{table*}

Fig. \ref{1206_histos} shows the distributions of $n$ on the left and $r_e$ on the right in each filter. 
The magenta lines denote the mean values of the respective distributions. 
Evidently, very many objects return the maximum output $n = $ 8 in the bluest filter, $B_{435}$. This constraint was likely met more often in this filter because of its low S/N ratio, which likely has caused Galapagos-2 to encounter difficulties in determining the correct intensity profiles for the respective galaxies. 
The distribution of $r_e$ of the sample galaxies shows no conspicuous features in any band, and the mean sizes of galaxies of $\sim 2 \mathrm{kpc}$ are within expectations. 

Table \ref{1206_n_re_var_table} lists the means of S\e rsic index $n$ and sizes in kpc for each band, as well as their uncertainties, estimated as $\sigma/\sqrt{N}$ (i.e., the standard error of the mean), where $\sigma$ is the standard deviation and $N$ is the number of galaxies. 
For our sample galaxies, $n$ is marginally dependent on wavelength, while galaxy sizes decrease with increasing wavelength, with $r_e$ being on average \s 20~\% smaller in near-infrared than in the optical bands.

To quantify the variation in effective radius and S\e rsic index in wavelength, we used the measurement ratios in the two bands $H_{160}$ and the $r_{625}$. 
We used the notation that was introduced by \citet{vulcani_14_gama_sizes_and_profiles_with_megamorph} and \citet{kennedy_15_gama_wavelength_dependence_structure}, $\mathcal{R}=r_{e(H_{160})}/r_{e(r_{625})}$ and $\mathcal{N}=n_{(H_{160})}/n_{(r_{625})}$. 
The selected filters provide robust values and cover a wide range of wavelengths, and were also used throughout the stellar population analysis in Sect. \ref{sec5}. 
Our tests show that including the low S/N filters $B_{435}$, $V_{606}$ , and $g_{475}$ leads to similar results, but we opt for the conservative approach of including only high-quality data.

In this context, $\mathcal{N} < 1$ corresponds to galaxies with a higher central light concentration in the blue $r_{625}$ band than in the red $H_{160}$ filter. 
$\mathcal{N} > 1$ conversely relates to the opposite behavior, and $\mathcal{N}$ \s\ 1 implies similar shapes of the light profiles in both passbands. 

The parameter $\mathcal{R}$ describes the variation in size in these two filters, with $\mathcal{R} < 1$ indicating a smaller size in the redder $H_{160}$ band. As a result, the centers appear redder, while the outskirts present themselves bluer in comparison. For $\mathcal{R} > 1$ the opposite is true, with the centers being bluer than the outskirts, whereas $\mathcal{R}$ \s\ 1 implies a similar size in both passbands. Hence, much like color gradients, $\mathcal{R}$ serves as an indicator of the color variation as a function of radius within an object.

With a mean of $\mathcal{N} = $ 0.92 $\pm$ 0.02 \footnote{ with the uncertainties calculated using the $\sigma/\sqrt{N}$ estimation}, it is reasonable to consider $\mathcal{N}$ constant for our sample. This confirms the notion that the elliptical galaxies in our sample are best described by a one-component intensity profile. 

The mean of $\mathcal{R} = $ 0.77 $\pm$ 0.02 corresponds to our finding that the effective radii of elliptical galaxies  are $\sim$ 25~\% smaller in $H_{160}$ than in $r_{625}$. 
This behavior already indicates that the majority of the studied galaxies possess redder centers and bluer outskirts, amounting to negative color gradients. Different stellar populations with similar light distribution, where blue, that is, younger, stars extend to larger radii can be a probable explanation, which we examine in Sect. \ref{sec5}.

\subsection{Color gradients}

To derive color gradients for every galaxy, we must first produce surface brightness profiles in each observed filter. They are based on the values of magnitudes, S\e rsic index $n,$ and effective radius $r_e$ as returned by the model fitting process. 
The S\e rsic profile is described as

\begin{equation}
\centering
I(R) = I_e \mathrm{exp} \left\{ -b_n \left[ \left( \frac{R}{r_e} \right)^{1/n} -1 \right] \right\},
\label{sersic_equation}
\end{equation}

where $I_e$ is the intensity at the effective radius $r_e$, which is the radius at which half of the total emitted light of the galaxy is encompassed. This in turn is ensured by the dimensionless analytically derived constant $b_n$. 
The S\e rsic index $n$ is a measure for the curvature of the profile. 
Following the derivations of \citet{graham_05_sersic_and_stuff} and using the results of an asymptotic expansion of $b_n$ given in \citet{ciotti_99_sersic_params}, we converted the S\e rsic profile to derive the surface brightness profile as

\begin{equation}
\centering
\mu (R) = \mu_e + \frac{2.5b_n}{\mathrm{ln}(10)}\left[ \left( \frac{R}{r_e} \right)^{1/n} -1 \right].
\label{sbp}
\end{equation}

\begin{figure*}
\centering
\includegraphics[width = \textwidth]{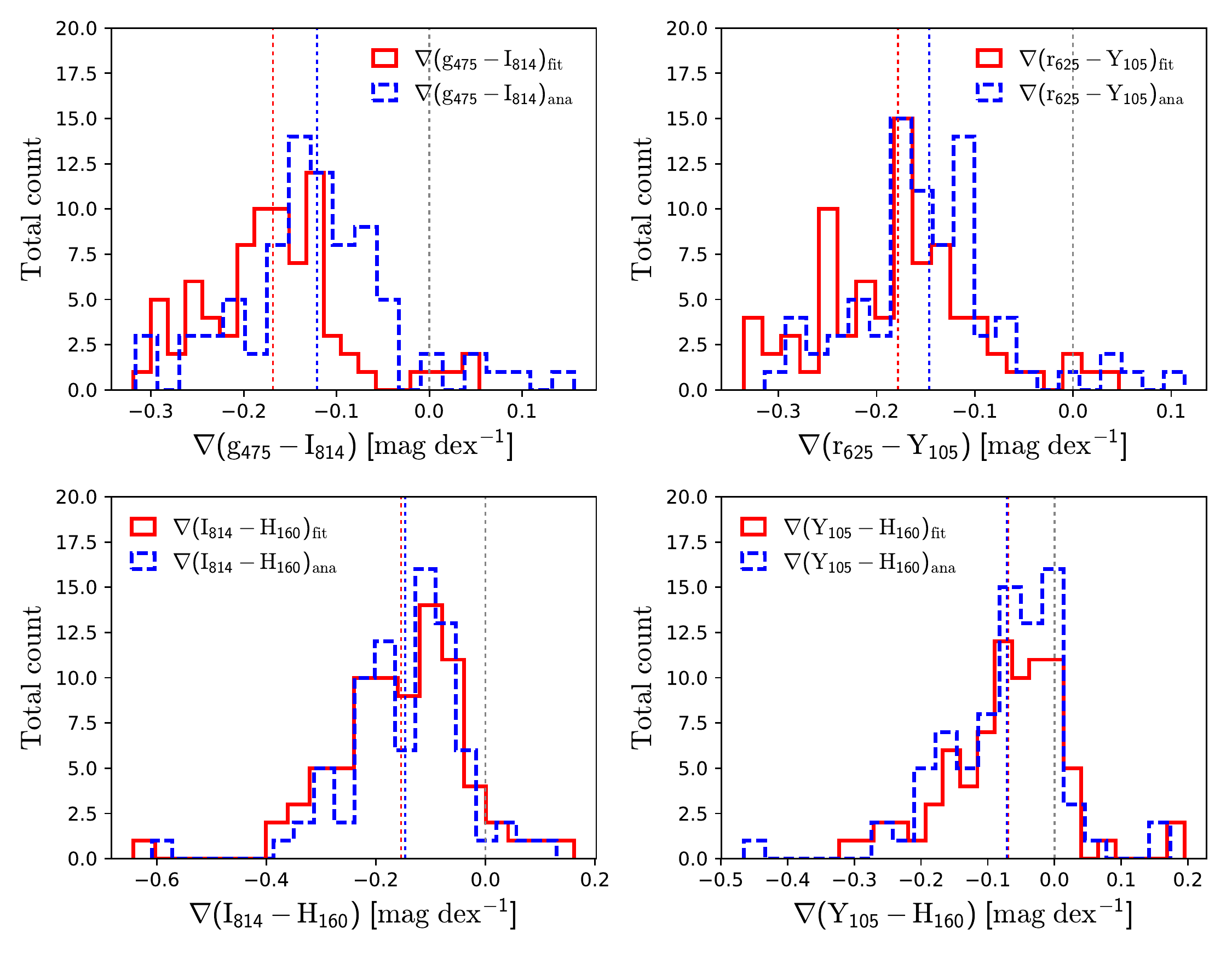}
\caption{Distributions of the color gradients for the four colors. Red bars indicate the color gradients assessed by fitting to the color profile, blue bars display the analytically derived gradients. The corresponding dashed lines denotes the respective means, and the gray dotted line indicates a color gradient $\nabla_{(\lambda_1 - \lambda_2)} = 0$. For all four colors and independent of  method, we derive negative mean color gradients.}
\label{1206_grad_histo}
\end{figure*}

$\mu_e$ depicts the surface brightness at a certain  $r_e$, that is, the mean effective surface brightness, which is described as

\begin{equation}
\langle \mu \rangle_e = \mu_e - \mathrm{2.5 log}\left(\frac{n\mathrm{e}^{b_n}}{b_n^{2n}}\Gamma(2n) \right),
\label{sersic_equation_3}
\end{equation}

where $\langle \mu \rangle_e$ is the mean surface brightness. This variable is expressed using the total apparent magnitude $m_{tot}$ and $r_e$:

\begin{equation}
m_{tot} = \langle \mu \rangle_e -2.5 \mathrm{log}(2\pi r_e^2).
\end{equation}

Because $n$ and $r_e$ generally vary from one filter $\lambda_1$ to another filter $\lambda_2$, the surface brightnesses differ in these two passbands. 
The color is then defined as the difference of two profiles at a certain radius, and as a consequence, the color profile is defined as the radial color trend of a galaxy. It is expressed as

\begin{gather}
\begin{aligned}
\left( \mu_{\lambda_1} - \mu_{\lambda_2} \right) (R) = \mu_{e,\lambda_1}+\frac{2.5b_{n,\lambda_1}}{\mathrm{ln}10} \left[ \left( \frac{R}{r_{e,\lambda_1}} \right)^{1/n_{\lambda_1}}-1 \right] \\ -  \mu_{e,\lambda_2}+\frac{2.5b_{n,\lambda_2}}{\mathrm{ln}10} \left[ \left( \frac{R}{r_{e,\lambda_2}} \right)^{1/n_{\lambda_2}}-1 \right]. \hspace{-0.15cm}
\end{aligned}
\label{color}
\end{gather}

Finally, the color gradient is defined as the slope of the color profile between a chosen radius interval, commonly scaled logarithmically:

\begin{equation}
\nabla_{\lambda_1 - \lambda_2} = \frac{\Delta \left( \left( \mu_{\lambda_1}-\mu_{\lambda_2}\right)(R) \right)}{\Delta \mathrm{log}R}.
\end{equation}

We used two approaches to derive color gradients. In the first, we applied a linear least-squares fit to the derived color profiles. In Fig. \ref{1206_grad_histo}, results for this approach are shown as red solid lines.
The second, analytic method uses the formula presented by \citet{la_barbera_02_opt_and_struct_prop}:

\begin{gather}
\begin{aligned}
\nabla_{\lambda_1 - \lambda_2} = \frac{2.5\mathrm{log}(\mathrm{e})}{ \mathrm{log}(r_M)-\mathrm{log}(r_m)}  \times \left\{b_{n,\lambda_2}\left[ \left( \frac{r_m}{r_{e,\lambda_2}}\right)^{1/n_{\lambda_2}}- \left( \frac{r_M}{r_{e,\lambda_2}}\right)^{1/n_{\lambda_2}}\right] \right. \hspace{-0.7cm}\\[3mm]  \left. +b_{n,\lambda_1}\left[ \left( \frac{r_M}{r_{e,\lambda_1}}\right)^{1/n_{\lambda_1}}- \left( \frac{r_m}{r_{e,\lambda_1}}\right)^{1/n_{\lambda_1}}\right] \right\}, \hspace{-0.1cm}
\end{aligned}
\label{la_barbera_grad}
\end{gather}

where the parameters $r_m$ and $r_M$ are the inner and outer limits in units of $r_e$ of the chosen radius interval and certain reference band to calculate the color gradients within. To be consistent with previous studies, we adopt $r_m$ = 0.1 $r_{e,625}$ and $r_M = r_{e,625}$. In Fig. \ref{1206_grad_histo} we present the results for this method as blue dashed lines.

The colors we examined are $g_{475} - I_{814}$, $r_{625} - Y_{105}$, $I_{814} - H_{160}$ , and $Y_{105} - H_{160}$, which at $z \sim $ 0.4 correspond to $U - V$, $B - R$, $V - Y,$ and $R - Y$ in rest-frame. 
The high quality and deep imaging of our HST data allows intervals of color gradients for our early-type sample to range from 0.1 - 2$r_e$. 

Table \ref{1206_grad_means} presents our results of derived color gradients as their means of all galaxies. For each of the four colors (Col. 1), we provide the means $\mu$ and standard deviation $\sigma$ derived from the linear fits (Cols. 2 and 4) and analytical formula (Cols. 3 and 5).

\begin{table}
\caption{Mean color gradients $\mu$ and the corresponding standard deviations $\sigma$ for each of our colors and assessment methods.}             
\label{1206_grad_means}      
\centering  

\begin{tabular}{ccccc}      
\hline\hline       
& $\mu_{fit}$ & $\sigma_{fit}$ & $\mu_{ana}$ & $\sigma_{ana}$ \\
& \multicolumn{4}{c}{[mag dex$^{-1}$]} \\
\hline
$\nabla_{g_{475}-I_{814}}$ & -0.17 & 0.08 & -0.12 & 0.09 \\ 
$\nabla_{r_{625}-Y_{105}}$ & -0.18 & 0.08 & -0.15 & 0.08 \\ 
$\nabla_{I_{814}-H_{160}}$ & -0.15 & 0.12 & -0.15 & 0.11  \\ 
$\nabla_{Y_{105}-H_{160}}$ & -0.07 & 0.09 & -0.07 & 0.09  \\
\hline
\end{tabular}
\end{table}

\begin{figure*}
\centering
\includegraphics[width = 20cm]{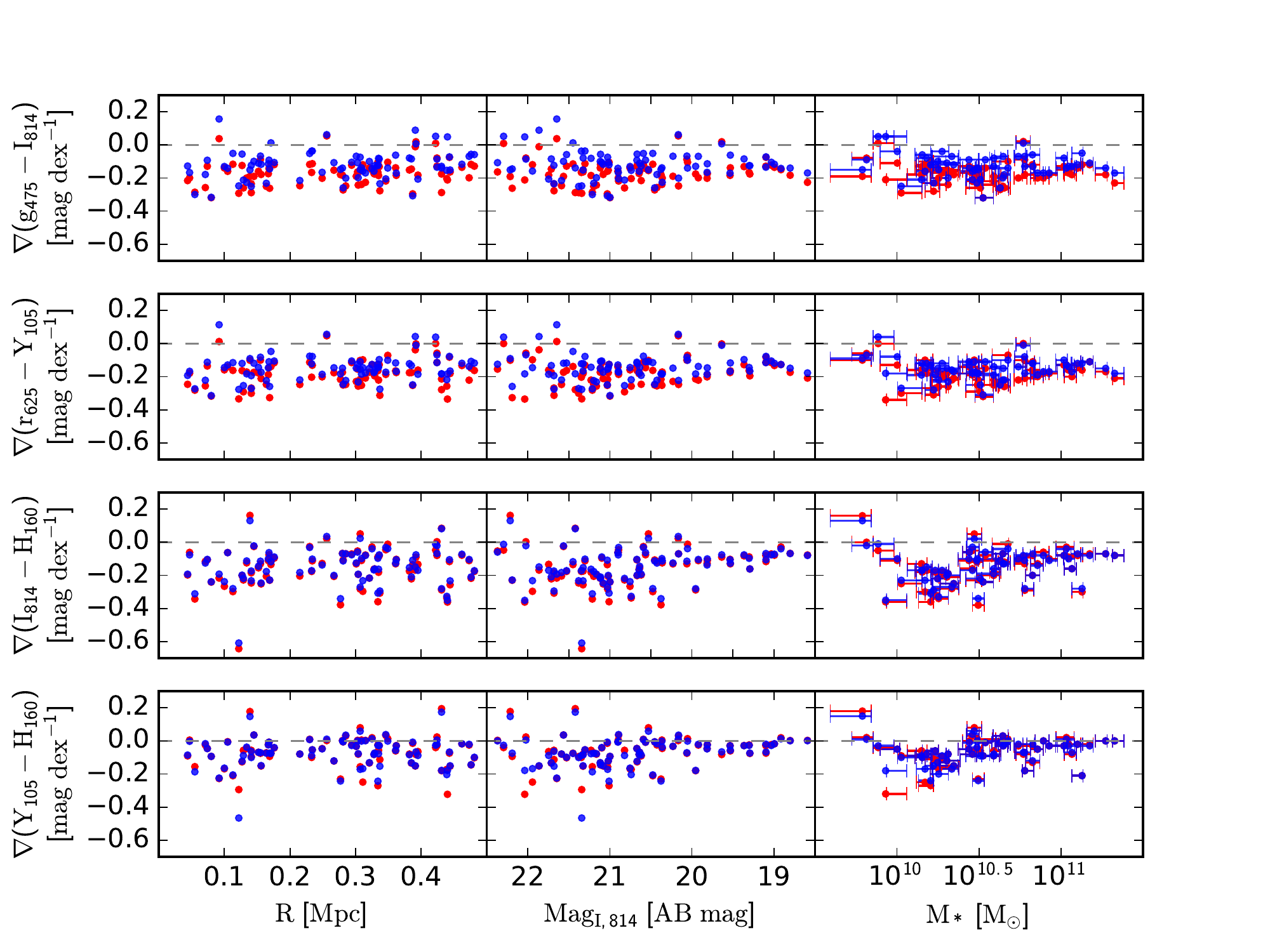}
\caption{Fitted and calculated color gradients of each individual early-type galaxy plotted against the respective cluster-centric distance (left), integrated magnitude in $I_{814}$ (middle) and stellar mass (right). The same color scheme as in Fig. \ref{1206_grad_histo} applies. The dashed, gray line  indicates a null color gradient. Based on Pearson correlation coefficients we detect no dependence between color gradients and those galactic properties, independent of color and method.}
\label{grad_depends}
\end{figure*}

Independent of the derivation method, all color gradients show a negative trend, that is, a decrease with radius. This means that the centers of galaxies are generally redder than their exterior regions, or, reversely, galaxies become bluer with radius. This observation refers to a variation in stellar populations at different radii in early-type galaxies in cluster centers. 
Color gradients become shallower the smaller, but also the redder the color interval. Therefore, the mean color gradients in $Y_{105}-H_{160}$ are rather shallow, as both filters trace a red and old population. The remaining colors also detect stars with intermediate ages and thus lead to steeper color gradients. The $g_{475}$ band lies blueward of the 4000 \angstrom ngstrom break at the cluster redshift, which would allow the detection of younger stellar populations and examination of their impact on the color variation with increasing radius. However, as the mean color gradient of $\nabla_{g_{475}-I_{814}}$ does not differ significantly from those of $\nabla_{r_{625}-Y_{105}}$ and $\nabla_{I_{814}-H_{160}}$ , we conclude that our sample galaxies consist on average predominantly of stars with intermediate and older ages $\geq 1-2 Gyr$.

We find slight discrepancies for $g_{475}-I_{814}$ and $r_{625}-Y_{105}$ , which we explain in two ways: First, the uncertainties in the fitting process are greater in the shorter wavelength $g_{475}$ than in $I_{814}$. This ultimately leads to greater inaccuracies of the intensity profiles, color profiles, and gradients.
Second, by using the parameters $r_m$ and $r_M$ , we introduce an additional error source. 

In Fig. \ref{1206_grad_histo} we show the distributions of color gradients for the four respective colors. Color gradients derived by the fitting method are shown as solid red lines, and gradients derived from the analytical approach are presented as blue dashed lines. We also indicate the corresponding means as vertical dotted lines and the position of a null color gradient ($\nabla_{(\lambda_1 - \lambda_2)} = 0$) as a dotted gray line.

In our sample, negative or neutral color gradients dominate, ranging primarily from \s\ 0.1 to \s\ -0.3 difference in magnitude per dex in radius. 
Differences between the two derivations (least-squares fit and analytical method) decrease with increasing wavelength. Results disagree most strongly in the $g_{475}-I_{814}$ color, which we explain by the low S/N in $g_{475}$.

In summary, we find that the 79 early-type galaxies of our sample generally have negative color gradients in $g_{475}-I_{814}$, $r_{625}-Y_{105}$, $I_{814} - H_{160}$, and $Y_{105} - H_{160}$. The behavior of some galaxies deviates, as we describe below.

\begin{itemize}
\renewcommand\labelitemi{--}
\item One galaxy (\s\ 1.3\% of this sample) has a color gradient $> 0.1$ mag dex$^{-1}$ when derived with the analytic approach in colors $g_{475}-I_{814}$ and $r_{625}-Y_{105}$. This galaxy is the smallest in our sample. It is likely that the size was underestimated during the fitting routine in the blue bands.
\item One galaxy (\s\ 1.3\% of this sample) has a color gradient $> 0.1$ mag dex$^{-1}$ using either derivation methods in $I_{814} - H_{160}$. This galaxy is a faint, low-mass galaxy that is surrounded by two bright companions. This could have led to an inaccurate size determination.
\item Two galaxies (\s\ 2.6\% of this sample) have a color gradient $> 0.1$ mag dex$^{-1}$ using either derivation methods in $Y_{105} - H_{160}$. One of these galaxies is the same as mentioned above in the results for $I_{814} - H_{160}$, the other does not show any distinct features that might explain a positive color gradient. However, with a photometric redshift difference of $\Delta z~\sim $~0.07, this source may not be a cluster member.
\end{itemize}

Additionally, we analyzed possible correlations between the derived color gradients and cluster-centric distance, integrated magnitude in $I_{814}$ , and stellar mass (Figure \ref{grad_depends}). We calculated the 24 Pearson correlation coefficients for each individual color-gradient measurement and galactic property. The maximum correlation coefficient is $\rho$ \s\ 0.22, and the minimum correlation coefficient amounts to $\rho$ \s\ -0.23. The mean correlation coefficients averaging over the methods to derive the color gradients and colors, that is, calculating the mean of the coefficients for the three galactic properties, are $\rho$ \s\ 0.14 for distance, $\rho$ \s\ -0.14 for integrated magnitude, and $\rho$ \s\ 0.08 for stellar mass.

\section{Comparison with simple stellar population models}
\label{sec5}

We compared the derived color gradients of our sample of 79 early-type galaxies to predictions by stellar population synthesis models to infer ages and metallicities of the underlying stellar populations. 
For this, we adopted singular stellar population (SSP) models using the tool \textit{EzGal} \citep{mancone_12_ezgal}.
This \texttt{Python} wrapper produces synthetic magnitudes and colors for galaxies depending on a given formation redshift $z_f$, metallicity $Z$, starformation history, and inital mass functions (IMF) and gives a choice of different stellar synthesis models, such as those published by \citet[BC03, ][]{bruzual_03_sps},  \citet{maraston_05_evo_pop_synth_models_analysis_...}, or \citet{conroy_09_ssp}. One advantage of this tool is that it allows redshifting the model spectra by convolving them with the filter curves and applying the corresponding $k$-corrections. This returns values in observed frame for the desired filters at a predetermined redshift. We therefore directly compare measured colors to a variation of modelled colors that depend on input ages and metallicities. In this way, we link observed parameters to formation conditions of galaxies, and ultimately constrain their evolutionary histories. 

For our comparison, we adopted a Chabrier IMF \citep{chabrier_03_IMF} and chose stellar population models by BC03 because of their range of available metallicities ($Z \sim [-0.4,\ 0.0,\ 0.4]\  $ [Log(Z/Z$_{\odot}$)] with Z$_{\odot}$ = $0.02$). Since these three metallicities are already implemented in \textit{EzGal}, it offers the option to calculate additional colors and magnitudes by interpolating additional models with other metallicities within the limits we described. 
In our selection process, we only selected elliptical, early-type galaxies for our analysis. 
In general, these are characterized by a singular star formation epoch at the formation redshift and a subsequent passive evolution. In this case, SSP models are sufficiently accurate for our purpose. We calculated models for 12 different formation redshifts between $z_f~\sim~[0.5,10]$ and five different metallicities in the interval $Z~=~[0.008,0.05]$.

Optical colors are affected by the age-metallicity degeneracy \citep{worthey_94_stellar_pop_models_disentanglement_age_Z}. Essentially, spectra of two unresolved galaxies are indistinguishable if one is three times older but possesses just half the metallicity of the other.
To break this degeneracy and to avoid any assumptions regarding the ages and metallicities of our model galaxies, we created a grid of two colors at our cluster redshift, whose predicted values depend on different SSPs based on varying ages and metallicities. 
This grid is similar to a grid spanned by the correct choice of Lick abundance indices and can therefore trace ages in one direction and metallicity in the other. 

Following \citet{smail_01_phot_study_ages_Z_etg_cluster} and \citet{la_barbera_02_opt_and_struct_prop}, a combination of optical and optical-infrared colors appears best suited to resolve the age-metallicity degeneracy. The first color traces the age of the stars, and the second is determined by their metallicity. For our grid, we therefore chose the color combination $r_{625} - Y_{105}$ for the optical part and $Y_{105} - H_{160}$ for the optical-IR part. These colors translate at the redshift $z \sim 0.4$ into $B - I$ and $I - Y$ in rest-frame. 

For each galaxy we then matched their colors, derived at 0.1, 0.5, 1, and 2 effective radii of the galaxy, to the model grid colors to infer values of ages and metallicities.
For our purpose, the flux at radii $>$ 2 $r_e$ becomes insignificant, and we find that light at these large radii does not contribute to the overall measurements of ages and metallicities.

\begin{figure}
\centering
\includegraphics[width = 9cm]{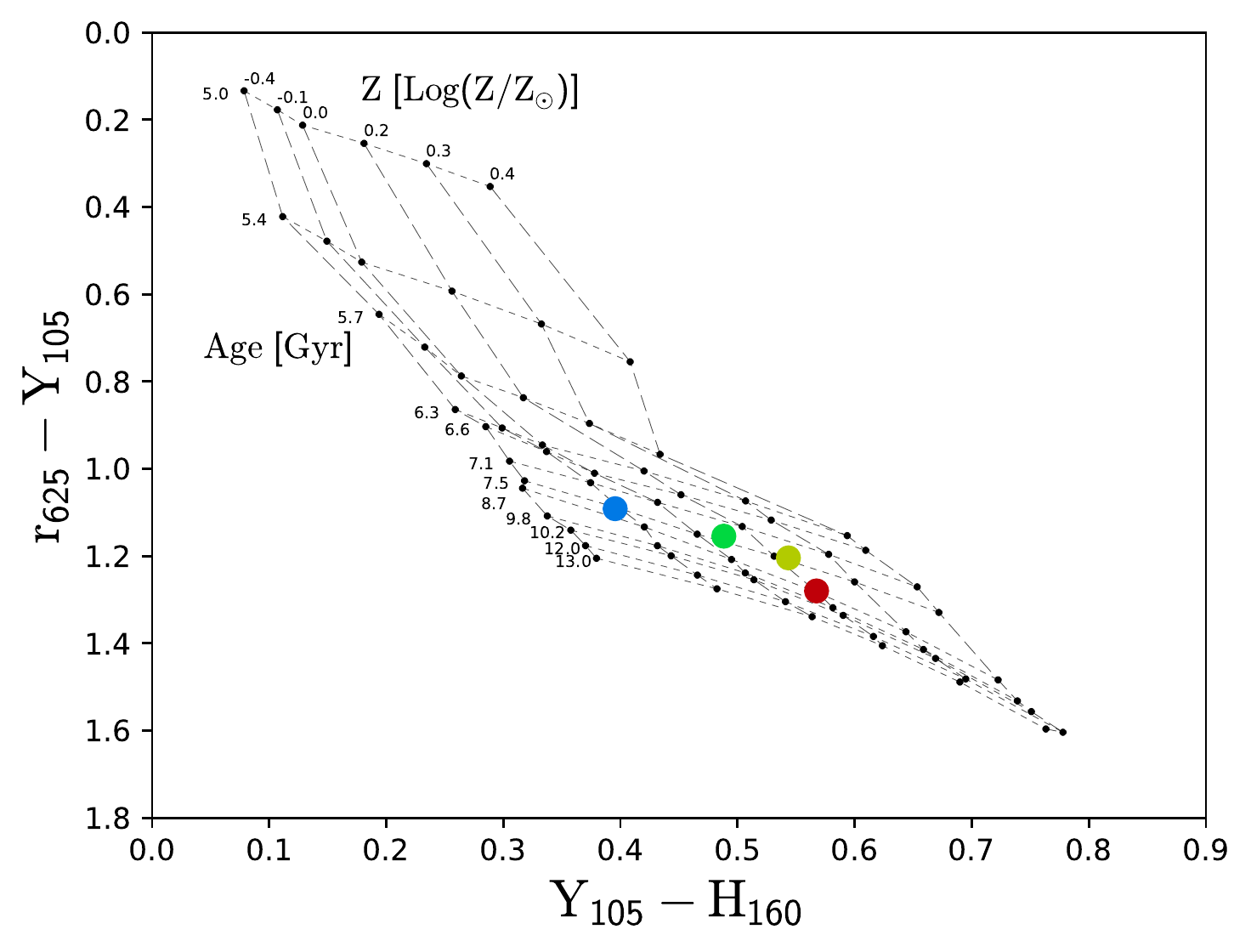}
\caption{Color-color diagram for one of the early-type cluster galaxies in our sample. The red dot represents the color values at 0.1$r_e$, the yellow dot shows the color values at 0.5$r_e$, the green dot  plots the color values at 1$r_e$ , and finally, the blue dot shows color values at 2$r_e$.  $r_{625} - Y_{105}$ is more sensitive to age variations, but $Y_{105} - H_{160}$ traces the metallicity. For this galaxy the innermost region between 0.1$r_e$ and 0.5$r_e$ is clearly age dominated, while at larger radii, a trend with metallicity takes over.}
\label{1206_87_radii_grid}
\end{figure}

We illustrate this approach in Fig. \ref{1206_87_radii_grid}, which shows the color values for one sample galaxy. The colored markers correspond to the galaxy color measurements at the four radii. 
In this example, we find that the central 0.5$r_e$ have a negative age gradient and constant metallicity. Between 0.5$r_e$ and 2$r_e$, stellar ages remain constant, and the  metallicity gradient becomes negative.
This shows that in order to explain the global appearance of the negative color gradients we detected in most of our galaxies, a combination of age and metallicity variation has to be taken into account.

A potential error source is that at solar metallicities, age tracks between 7 and 9 Gyr are almost identical. However, because we are interested in the relative differences of the ages and metallicities, we can neglect this problem.
Similarly, we omit the impact of dust, which can be regarded as a minor contribution in early-type galaxies.

\begin{figure}
\centering
\includegraphics[width = 9cm]{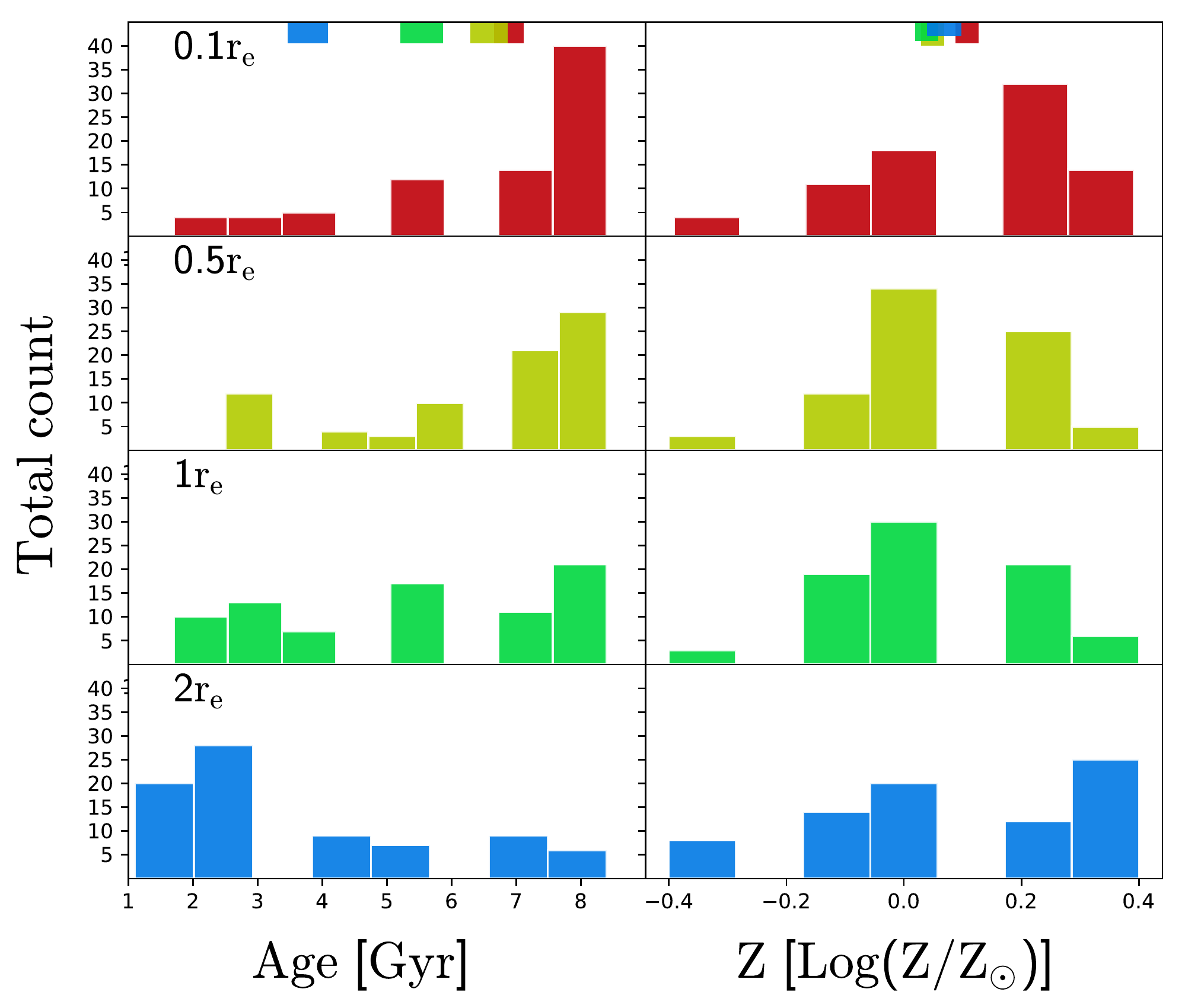}
\caption{We show in the left panels from top to bottom the age distributions from 0.1$r_e$ to 2$r_e$ . Analogously, the right panels show the metallicity distributions. The color scheme is the same as in Fig. \ref{1206_87_radii_grid}. The rectangles at the top display the average values, and the widths correspond to the respective errors. For our sample galaxies we detect on average a negative metallicity gradient (i.e., higher metallicities in the centers) and a clear negative age gradient (i.e., older stars in the central regions).} 
\label{1206_age_Z_histos}
\end{figure}

We applied the same analysis to all 79 galaxies of our sample. We then considered the relative age and metallicity variations with radius. Fig. \ref{1206_age_Z_histos} summarizes results for derived ages on the left and metallicities on the right at the four galaxy radii 0.1, 0.5, 1, and 2$r_e$.
To this end, we divided our age range into eight bins of equal size from 1 Gyr to 9 Gyr. 
Stars in galaxy centers (within 0.1$r_e$) are the oldest. Half of the sample galaxies have central stars that are 8-9 Gyr old (which is the maximum attainable age). 
Stellar populations are younger at larger radii. At 0.5 and 1$r_e$, the mean stellar ages are 6.58$\pm$0.29 Gyr and 5.54$\pm$0.33 Gyr, respectively.
Only \s\ 6\% of the sample have old stellar populations at 2$r_e$. The majority (\s 60\%) have stars that are 1-3 Gyr old at this radius.
This translates into a mean age difference between the central regions and the outskirts of \s 3 Gyr. 

Metallicity gradients are negative in most galaxies (right panel of Fig. \ref{1206_age_Z_histos}). 
In the galaxy centers, at 0.1$r_e$, we determine  0.11$\pm$0.02 [Log(Z/Z$_{\odot}$)] to supersolar values. At radii 0.5, 1, and 2$r_e$ from the center, we measure solar metallicities, with 0.05$\pm$0.02[Log(Z/Z$_{\odot}$)], 0.04$\pm$0.02[Log(Z/Z$_{\odot}$)], and 0.07$\pm$0.03[Log(Z/Z$_{\odot}$)], respectively. 

\section{Discussion}
\label{sec6}

The characteristics revealed in our analysis of color, age, and metallicity gradients of the early-type population of galaxies in the core of the cluster MACS 1206 at z$\sim$0.44 agree well with the currently accepted scientific consensus established by previous studies.
Our measurements of 79 sample galaxies in 12 HST filters reveal that early-type galaxies are 25\% smaller in infrared ($H_{160}$) than in optical ($r_{625}$) filters, while the S\e rsic index $n$ is constant within 3$\sigma$ over this range.
The mean S\e rsic index $\langle n \rangle$ of our sample over the entire wavelength range is $n \sim$ 4.59 $\pm$ 0.04.
This is consistent with measurements from \citet{vulcani_14_gama_sizes_and_profiles_with_megamorph}, who found a constant $\langle n \rangle \sim $ 4 over all filters used in the \textit{GAMA} survey, from the $g$ band and extended down to the $K$ band. Additionally, they also detected a decrease of \s~40\%~ for $r_e$ going from $u$ to $H$. Similar trends have been found by \citet{kelvin_12_gama_struc_investigation_via_model}, who established a 38\%\ decrease in $r_e$ for the same sample as well, but used only single-component fits. 
Finally, \citet{kennedy_15_gama_wavelength_dependence_structure} confirmed these results in their investigation of \textit{GAMA II} field galaxies.
On the other hand, \citet{la_barbera_10_gal_params_grizYJHK} found higher medians ($\langle n \rangle$ \s\ 6) by examining local galaxies. However, their wider wavelength range covered filters from the $g$ band to the $K$ band, which could explain some of the discrepancies. 
They also found a 35\% larger decrease in sizes. 

For a sample at intermediate redshifts in clusters, which is comparable to ours, \citet{la_barbera_02_opt_and_struct_prop, la_barbera_03_evo_UV_NIR_struc_props_cluster_gal} found a similar behavior of the dependence of $r_e$ and $n$ on wavelength.
For their investigation, they examined approximately 270 galaxies in three different clusters between $z = $ 0.21 and $z = $ 0.64 in $U, V,$ and $H$ rest-frame filters. Their results,
$r_{e,UV}/r_{e,opt} = 1.2 \pm 0.05$, $r_{e,opt}/r_{e,IR} = 1.26 \pm 0.06,$ and $n_{UV}/n_{opt} = 1.0 \pm 0.1$ and $n_{opt}/n_{IR} = 0.88 \pm 0.03$ are in good agreement with our assessment and are consistent with the local values.
Even in clusters at $z$ = 1.39, passive galaxies still show the same trends, as shown by \citet{chan_16_sizes_colorgrads_etc_in_massive_cluster}, who derived a \s~20\% decrease in galaxy size from $i_{775}$ to $H_{160}$.
 
We explain the dependence of galaxy sizes on wavelength with intrinsic color gradients in stellar populations of early-type galaxies. The majority of early-type galaxies have red centers and in comparison bluer regions at larger radii from the center. Since we assumed a negligible amount of dust in early-type galaxies, we interpret this as a result of older stellar populations residing in the centers of our sample galaxies. 

The medians of color gradients in our sample galaxies in four different colors range between -0.07 $\pm$ 0.02 and -0.18 $\pm$ 0.03 mag dex$^{-1}$.
The color gradients we determined in our sample are consistent with the findings of \citet{saglia_00_evolution_color_gradients_of_etg}, who derived color gradients for $z$ \s\ 0.4 cluster galaxies. In the rest-frame colors $U-B$, $U-V$ and $B-V$, which are similar to the colors presented in this work, all their objects have a negative gradient in at least one color.
Other studies treating galaxies at similar redshifts are \citet{tamura_00_color_gradients_etg}, who find a median gradient of $\nabla (\lambda_{B,450} - \lambda_{I,814}) = -0.23 \pm 0.05$ mag dex$^{-1}$ , or  \citet{la_barbera_02_opt_and_struct_prop}, who likewise derived comparable values for color gradients in elliptical galaxies that agree well with our determinations. Furthermore, our assessments are consistent with the mean color gradients obtained by \citet{la_barbera_03_evo_UV_NIR_struc_props_cluster_gal} with $\nabla (\lambda_{UV} - \lambda_{opt}) = -0.18 \pm 0.04$  mag dex$^{-1}$ and $\nabla (\lambda_{opt} - \lambda_{IR}) = -0.4 \pm 0.1$ mag dex$^{-1}$.
In an early work by \citet{peletier_90_ccd_surface_photometry_of_galaxies,peletier_90_nir_photometry_etg}, similar results were found for local early-type galaxies. They measured mean color gradients of -0.20 $\pm$ 0.02 mag dex$^{-1}$ in $U-R$ and -0.09 $\pm$ 0.02 in $B-R$. 

Studies at higher redshift elliptical galaxies, however, reveal steeper negative color gradients than those established in local and intermediate-$z$ galaxies. \citet{gargiulo_12_colour_and_stellar_prop_etg} found gradients between -0.1 $\pm$ 0.1 and -1.0 $\pm$ 0.1 mag dex$^{-1}$ for their sample of early-type galaxies at 1.0 $< z <$ 1.9.  \citet{guo_11_color_and_stellar_pop_gradients} also reported steeper color gradients for their $z$ \s\ 2 galaxies, as did \citet{de_popris_15_morphological_evo_red_sequence}, who derived a median color gradient of -0.25  mag dex$^{-1}$ for their high-redshift ($\langle z \rangle$ \s\ 1.25) selection. Finally, in their investigation of cluster galaxies at $z$ \s\ 1.39, \citet{chan_16_sizes_colorgrads_etc_in_massive_cluster} found a median value of \s\ 0.45 mag dex$^{-1}$, which is a factor of two steeper than the local values. 

We have shown that color gradients are the result of a combination of age and metallicity gradients. 
We derive metallicity gradients of \s\ -0.2 dex per decade in radius, which confirms previous results found by \citet{saglia_00_evolution_color_gradients_of_etg}, \citet{la_barbera_03_evo_UV_NIR_struc_props_cluster_gal}, and \citet{tamura_00_origin_color_gradients_etg}. Stellar populations in the internal regions of our sample galaxies are on average \s3 Gyr older than those in the external regions. This result is slightly more extreme, but within the error estimations of the same studies. 

The continuing inflow of metal-rich gas means that a monolithic collapse scenario requires much steeper negative metallicity gradients (between -0.5 and -1 dex per radial decade) than those observed. 
Consequently, the monolithic collapse theory also predicts a positive age gradient: As a result of gas enrichment, the younger stellar population would need to be located in the innermost regions of the galaxy, and the stellar ages would increase with radius. 
Our observations show, however, that elliptical galaxies have negative age gradients.

Gas-rich mergers are unlikely to play an important role in the recent evolution ($z \lesssim 0.5 - 1$) of the early-type population we investigated. A recent merger history would predict an old, red, and extended component superimposed by a more concentrated young and blue stellar population. 
Frequent active mergers lead to differences in mean S\e rsic indices and sizes in local galaxies compared to those at intermediate redshifts. This has not been confirmed.

Our results support the current evolution scenario, where a series of dry mergers contributes to an inside-out growth of early-type galaxies. 
During this process, young and metal-poor stars of low-mass galaxies are accreted by high-mass galaxies and build an extended component.
This leads to a significant evolution in size, with little increase in mass \citep{driver_13_Two_phase_galaxy_evo_cosmic_SFH, naab_09_minor_merger_and_size_evo_etgs, oser_two_phase_formation}.  
This scenario matches observations of very compact, massive quiescent galaxies at $z$ \s\ 2 well \citep[][and references therein]{davari_14_how_robust_size_measurements_galfit}, which are thought to be the progenitors of early-type galaxies at low and intermediate redshifts.
The inside-out growth through dry mergers also dilutes age and metallicity variations, which matches our observations of gradients as well as observations at higher and lower redshifts.

The cluster galaxies of our sample are located in the central region of a massive galaxy cluster (Fig. \ref{1206_selection}). During our selection processes, we further ensured that our sample is comprised of red-sequence galaxies that have smooth and spheroid-dominated morphologies. We are therefore confident to target the early-type population that is located well within the cluster. 
Within this central region, we found no correlation of the color gradients with cluster-centric distance, magnitude, or stellar mass. This in turn means that at least within the small area we investigated, the variation in stellar populations is independent of the environment, luminosity, or stellar mass. We conclude that the early-type galaxies, which are situated within the innermost region of the cluster and possess similar mass, undergo a similar evolutionary path, resulting in the observed lack of any correlation. 
In their investigation, \citet{saglia_00_evolution_color_gradients_of_etg} found similar conclusions with respect to galaxy luminosities, and \citet{guo_11_color_and_stellar_pop_gradients} and \citet{gargiulo_12_colour_and_stellar_prop_etg} confirmed the independence on stellar mass. Furthermore, \citet{goddard_17_spec_gradient_environment} found no correlation with environment either.

Galaxies in the central regions have formed at the earliest times of cluster assembly, and are therefore considered the oldest members of the cluster. 
Their evolutionary paths are expected to be similar, and our result supports this notion. 
In order to investigate the dependence of the cluster environment on stellar populations of its members, we will therefore have to expand the search to greater cluster-centric distances out to and beyond the virial radius of the cluster and also to lower stellar masses. In this way, we are able to follow the assembly history of the cluster and trace the effect of structure growth on galaxy population.

Galaxies falling into a cluster are affected in a number of ways. They experience physical processes that disrupt the gas supply and ultimately stop their star formation. In clusters, we observe a variety of transitional objects, such as passive spirals or post-starburst galaxies, whose stellar populations and morphologies indicate a recent change in star formation and/or a dynamical history. Cluster mechanisms that strip away the gas of galaxy halos and disks will dim and redden the galaxies as their stars age \citep{kuchner_17_paper}. These galaxies may be an important galaxy transformation pathway. An extension of our investigation of color gradients to galaxies located in cluster outskirts and infall regions of clusters will allow an exploration of environmental disk fading. In future publications, we plan to include the remaining CLASH clusters, as well as all galaxies, independent of morphological type.

\section{Summary}
\label{sec7}

We analyzed a total of 79 early-type galaxies, located in the central regions of the massive cluster MACS 1206 at $z$ = 0.44. We used HST imaging from the CLASH survey in 12 filters ranging from $B_{435}$ to $H_{160}$ to accurately determine the structural parameters. Membership determination is partially ensured through the additional spectroscopic information from the follow-up program CLASH-VLT and photometric redshifts calculated from 16 filters from UV to near-infrared.

We employed a 2D model fitting for each galaxy by using 12 HST filters simultaneously, using the tool Galapagos-2, which was produced by the \texttt{MegaMorph} project. 
In a visual inspection we only considered smooth and regular galaxies with S\e rsic indices $n >$ 2.5, which resulted in our final sample of 79 elliptical galaxies.
We used the structural parameters S\e rsic index $n$, effective radius $r_e$, and total magnitude $M_{tot}$ provided by the models to derive surface brightness profiles. These were used to produce radial profiles for the colors $g_{475} - I_{814}$, $r_{625} - Y_{105}$, $I_{814} - H_{160}$ , and $Y_{105} - H_{160}$.
Color gradients were derived as the logarithmic slopes of these color profiles by 1) approximating a least-squares fit, and 2) calculating the ratios of the effective radii in the two bands analytically.

We analyzed age and metallicity contributions to the color gradients by employing BC03 single stellar population models, adopting a Chabrier IMF. 
We allowed initial conditions of the models to vary in different formation redshifts from $z_f  = 0.5$ to $z_f = 10$, and in metallicities, where we limited the values to lie between $Z = 0.05$ and $Z = 0.008$. 
Our relative determinations of estimated ages and metallicities are consistent with those of previous studies. Our comparison of simulated colors at the cluster redshift with observed colors constrains ages and metallicities of our sample galaxies. We summarize our findings below.

\begin{itemize}
\renewcommand\labelitemi{--}
\item The effective radius decreases with increasing wavelength as $r_{e,160} \sim 0.75 r_{e,625}$. This is due to the variations of stellar populations in the galaxies and characterizes negative color gradients.
\item On average, the mean S\e rsic index $n$ appears constant over wavelength. This shows that early-type galaxies can indeed be described by a single-component intensity profile. 
\item The majority of the galaxies and the overall mean show negative color gradients of between -0.07 $\pm$ 0.02 and -0.18 $\pm$ 0.03 mag dex$^{-1}$, indicating a redder, older stellar population in the centers and bluer, younger stars in the outskirts. The results are independent of the derivation method and also differ insignificantly between the employed colors.
\item No correlation or dependence of the color gradients on stellar mass, magnitude, or cluster-centric distance can be found, implying that none of these properties are directly responsible for the radial color variations.
\item Age as well as metallicity drive the color gradients. While the centers appear to be \s~3 Gyr older than the exterior regions, the metallicity drops from slightly supersolar values in the innermost regions to solar values in the outskirts.
\item Considering all these observations, the most probable evolutionary scenario for elliptical galaxies is that they already assemble a majority of their stellar mass at very high redshifts ($z \gtrsim 2$), followed by a passive evolution of the initial stellar population. With time, the galaxies undergo several dry and minor mergers, resulting in a much more notable increase in size than in mass, as the more metal-poor and younger populations of the accreted galaxies settle into more extended structures. 

\end{itemize}

The examination of variations of colors within a galaxy, independent of morphology and especially redshift, can give valuable insights into its history. In particular with the start of the James Webb Space Telescope\texttt{,} it will be possible to expand this approach to more distant and fainter galaxies, which will allow for a more comprehensive picture of galaxy evolution.

\begin{acknowledgements}
      This study is based on data collected as part of the CLASH Multi-Cycle Treasury Program with the Hubble Space Telescope, as well as on data collected at the ESO-VLT (prog. ID 186.A-0798).
      This analysis relies on the work done within the CLASH and CLASH-VLT collaborations, who prepared the imaging and spectroscopic data. We thank the referee for the useful comments that improved our publication. The paper is based on the Master's thesis of VM, who thanks the University of Vienna for the grant near completion of studies.  
      This publication is supported by the Austrian Science Fund (FWF).
\end{acknowledgements}

\bibliographystyle{aa}
\bibliography{references}

\end{document}